\documentclass[12pt]{article}
\usepackage{amssymb}
\usepackage{latexsym}
\usepackage{amsmath}
\usepackage{graphicx}
\usepackage{amsthm}
\begin{document}
\title{Overall Dynamic Properties of 3-D periodic elastic composites}
\author{Ankit Srivastava \and Sia Nemat-Nasser\\
Department of Mechanical and Aerospace Engineering\\ 
University of California, San Diego\\
La Jolla, CA, 92093-0416 USA}
\maketitle
\abstract{A method for the homogenization of 3-D periodic elastic composites is presented. It allows for the evaluation of the averaged overall frequency dependent dynamic material constitutive tensors relating the averaged dynamic field variable tensors of velocity, strain, stress, and linear momentum. The formulation is based on micromechanical modeling of a representative unit cell of a composite proposed by Nemat-Nasser \& Hori (1993), Nemat-Nasser \emph{et. al.} (1982) and Mura (1987) and is the 3-D generalization of the 1-D elastodynamic homogenization scheme presented by Nemat-Nasser \& Srivastava (2011) .

We show that for 3-D periodic composites the overall compliance (stiffness) tensor is hermitian, irrespective of whether the corresponding unit cell is geometrically or materially symmetric.Overall mass density is shown to be a tensor and, like the overall compliance tensor, always hermitian. The average strain and linear momentum tensors are, however, coupled and the coupling tensors are shown to be each others' hermitian transpose. Finally we present a numerical example of a 3-D periodic composite composed of elastic cubes periodically distributed in an elastic matrix. The presented results corroborate the predictions of the theoretical treatment.}

\section{Introduction}

Recent interest in the character of the overall dynamic properties of composites with tailored microstructure necessitates a systematic homogenization procedure to express the dynamic response of an elastic composite in terms of its average effective compliance and density.

In the present paper we propose a method of homogenization of 3-D microstructured elastic composites which directly provides the overall frequency dependent dynamic material parameters. This method is the 3-D generalization of the 1-D homogenization scheme presented by Nemat-Nasser \& Srivastava (2011). The 1-D homogenization scheme was shown to be equivalent to the field variable integration scheme of Nemat-Nasser \emph{et. al.} (2010) in the limit of Bloch wave propagation. The present method doesn't require the pointwise calculation of the values of the field variables within the unit cell. Rather, it gives the solution of the elastodynamic field equation to any desired degree of accuracy. The method is inspired by the micromechanical homogenization scheme originally proposed by Nemat-Nasser \& Taya (1981) and further developed in Nemat-Nasser \emph{et. al.} (1982). A similar homogenization method has been used by Amirkhizi and Nemat-Nasser to calculate the effective electromagnetic properties of a periodic composite Amirkhizi \& Nemat-Nasser (2007).

In what follows we outline our homogenization approach for a general 3-D periodic elastic composite. We show that the resulting averaged dynamic constitutive parameters are tensorial in nature and that the average strain tensor is coupled with the average momentum tensor. Such a form of the averaged constitutive relation where the constitutive parameters (including mass density) are tensors and where the average strain (average stress) is coupled with average linear momentum has been predicted by Milton \& Willis (2007) and Willis (2009). See also Shuvalov \emph{et. al.} (2009). Willis has also shown that for elastodynamic problems with inelastic strain, the effective properties, thus calculated, are uniquely determined. See Willis (2011). We show that this form of the constitutive relation is the natural outcome of our homogenization method. Furthermore, we prove, on mathematical grounds, that the coupling parameters arising in the averaged constitutive relations are hermitian transpose of each other, irrespective of the material or geometric asymmetries.

\section{Microstructural homogenization of periodic composites}

Here we present a homogenization method based on micromechanical consideration of the volume averages of the field variables, viewed as measurable macroscopic physical quantities. We express the solution to the elasto-dynamic equations of motion as the sum of the volume average and a perturbation due to the heterogeneous composition of the unit cell,

\begin{equation}
\mathbf{Q}=\mathbf{Q}^0+\mathbf{Q}^p
\end{equation}
where $\mathbf{Q}$ represents the field variables, stress ($\boldsymbol{\sigma}$) or velocity ($\boldsymbol{\dot{u}}$).
The aim is to derive a set of constitutive relations for the overall averaged parts of the field variables, using the local elasto-dynamic equations of motion and constitutive relations.  
This then provides the homogenized frequency-dependent material parameters.  In what follows, we describe our approach using a periodic composite, but the final constitutive relations also apply  to a finite unit cell.

Consider harmonic waves in an unbounded elastic composite consisting of a collection of bonded, identical unit cells ($\Omega=\{x_i:-a_i/2\leq x_i<a_i/2;\; i=1,2,3\}$) which repeat themselves in all directions, and hence constitute a periodic structure. In view of the periodicity of the composite, we have $\rho(\mathbf{x})=\rho(\mathbf{x}+m'\mathbf{I}_\beta)$ and $\mathbf{C}(\mathbf{x})=\mathbf{C}(\mathbf{x}+m'\mathbf{I}_\beta)$; here $m'$ is an integer, $\rho(\mathbf{x})$ is the density, $\mathbf{C}(\mathbf{x})$ is the fourth order tensor of modulus of elasticity whose inverse is the compliance tensor $\mathbf{D}(\mathbf{x})$, and $\mathbf{I}_\beta,\beta=1,2,3$ denote the three vectors which form a parallelepiped enclosing the periodic unit cell. For time harmonic waves with frequency $\omega$, the field quantities are proportional to $e^{\pm i\omega t}$. For waves with wave vector $\mathbf{q}=q_i\hat{\mathbf{x}}_i$ where $\hat{\mathbf{x}}_i$ is the unit vector in the $i^\mathrm{th}$ direction and einstein summation convention applies, the Bloch representation of the field variables takes the form,

\begin{equation}\label{EBloch}
{\hat{\mathbf{Q}}(\mathbf{x},t)}=\mathrm{Re}\left[\mathbf{Q}(\mathbf{x})\exp[i(\mathbf{q}\cdot \mathbf{x}-\omega t)]\right]
\end{equation}
where $\hat{\mathbf{Q}}$ represents the field variables, stress ($\hat{\boldsymbol{\sigma}}$), strain ($\hat{\boldsymbol{\varepsilon}}$), momentum ($\hat{\mathbf{p}}$) or velocity ($\hat{\mathbf{\dot{u}}}$) whereas $\mathbf{Q}$ represents their periodic parts (${\boldsymbol{\sigma}}$, ${\boldsymbol{\varepsilon}}$, ${\mathbf{p}}$, ${\mathbf{\dot{u}}}$). The representation, Eq. \ref{EBloch},  separates the time harmonic and macroscopic factor from the microscopic part of the field variables.  Even for a finite unit cell, the Fourier series solution of the microscopic part is periodic so that this solution satisfies, $\mathbf{Q}(\mathbf{x})=\mathbf{Q}(\mathbf{x}+m'\mathbf{I}_\beta)$. We emphasize here that the frequency and the wavenumber, $\omega$ and $q$, are, at this point, unrelated and arbitrary. 

The local conservation and kinematic relations are,

\begin{equation}
\begin{array}{l}
\displaystyle \tilde{\nabla}\cdot\boldsymbol{\sigma}=-i\omega\mathbf{p}\\
\displaystyle (\tilde{\nabla}\otimes\mathbf{\dot{u}})_{sym}=-i\omega\boldsymbol{{\varepsilon}}\\
\end{array} 
\label{EFieldEqn}
\end{equation}
where $\tilde{\nabla}\rightarrow\nabla+i\mathbf{q} $.  
The corresponding constitutive relations are,

\begin{equation}
\begin{array}{l}
\displaystyle\boldsymbol{\sigma}= \mathbf{C}:\boldsymbol{\varepsilon}\\
\displaystyle \mathbf{p}=\rho\mathbf{\dot{u}}\\
\end{array} 
\label{EFieldEqn1}
\end{equation}
where $\mathbf{C(x)}$ is the tensor of elasticity and $\rho(\mathbf{x})$ is the mass-density.  These local material parameters represent the structure and composition of the unit cell.

Now we replace the heterogeneous unit cell with a homogeneous one having uniform density $\rho^0$ and stiffness $\mathbf{C}^0$. 
In order to reproduce the strain and momentum of the actual unit cell, field variables eigenstress (${\boldsymbol{\Sigma}}$) and eigenvelocities (${\dot{\mathbf{{U}}}}$), are introduced.  These quantities are then calculated, using the basic local field equations and constitutive relations.
The idea stems from the polarization stress or strain that was originally proposed by Hashin (1959) and further developed by Hashin \& Shtriktman (1962,1963) and later by others, in order to construct energy-based bounds for the composite's overall elastic moduli. 
The basic tool in these works has been the result, obtained by Eshelby (1957) in three dimensions and earlier by Hardiman (1954) in two dimensions, 
that the stress and strain are constant within an ellipsoidal (elliptical in two dimensions) region of an infinitely extended uniform elastic medium when that region undergoes a uniform transformation corresponding to a uniform inelastic strain. 

Here, however, we present a different tool that can be used to actually calculate the point-wise values of the elasto-dynamic field variables to any desired degree of accuracy.  
For this, we require that the actual values of the field variables at every point within the homogenized and the original heterogeneous unit cell be exactly the same.  
To ensure this, we require that the following consistency conditions hold at every point within the unit cell:
\begin{equation}
\begin{array}{l}
\displaystyle \mathbf{C}^0:\boldsymbol{\varepsilon}=\boldsymbol{\sigma}-{\boldsymbol{\Sigma}}\\
\displaystyle \mathbf{p}=\rho^0(\mathbf{\dot{u}}-{\dot{\mathbf{{U}}}})\\
\end{array} 
\label{Eeigenstuff}
\end{equation}

The eigenstress and eigenvelocity fields are zero in regions where the material properties of the heterogeneous unit cell are equal to the chosen uniform material properties, 
$\mathbf{C}^0$ and $\rho^0$. 
From Eqs. (\ref{EFieldEqn}, \ref{Eeigenstuff}) we have,

\begin{equation}\label{Ee2}
\tilde{\nabla}\cdot\mathbf{C}^0:(\tilde{\nabla}\otimes \mathbf{\dot{u}})_{sym}+\omega^2\rho^0\mathbf{\dot{u}}=\omega^2\rho^0\mathbf{\dot{U}}+i\omega(\tilde{\nabla}\cdot\boldsymbol{{\Sigma}})
\end{equation}

\begin{equation}\label{Ee1}
\mathbf{C}^0:[\tilde{\nabla}\otimes(\tilde{\nabla}\cdot\boldsymbol{\sigma})]_{sym}+\omega^2\rho^0\boldsymbol{\sigma}=\omega^2\rho^0\boldsymbol{\Sigma}+i\omega\rho^0\mathbf{C}^0:(\tilde{\nabla}\otimes\mathbf{\dot{U}})_{sym}
\end{equation}

Since the stress and displacement fields ($\mathbf{Q}$) are periodic with the unit cell, they can be expanded in a spatial Fourier series, 

\begin{equation}
\mathbf{Q}(\mathbf{x})=\mathbf{Q}^0+\mathbf{Q}^p=\langle \mathbf{Q}\rangle_\Omega+\sum_{\boldsymbol{\xi}\neq 0}\mathbf{Q}(\boldsymbol{\xi})\mathrm{e}^{i\boldsymbol{\xi}\cdot\mathbf{x}}
\end{equation}

\begin{equation}\label{EF}
\langle \mathbf{Q}\rangle_\Omega=\frac{1}{\Omega}\int_\Omega \mathbf{Q}(\mathbf{x})dV_x
\end{equation}

\begin{equation}
\mathbf{Q}(\boldsymbol{\xi})=\frac{1}{\Omega}\int_\Omega \mathbf{Q}(\mathbf{x})\mathrm{e}^{-i\boldsymbol{\xi}\cdot \mathbf{x}} dx
\end{equation}

\begin{equation}
\Omega=8a_1a_2a_3
\end{equation}

\begin{equation}
\boldsymbol{\xi}=\xi_i\hat{\mathbf{x}}_i;\quad \xi_\alpha=n_\alpha\pi/a_\alpha;\quad n_\alpha \;\mathrm{integers}
\end{equation}
where greek indices are not summed. $\langle \mathbf{Q}\rangle_\Omega$ represents the averaged value of the field variable over the unit cell and appears in its macroscopic description, and $\mathbf{Q}(\boldsymbol{\xi})$ represents the local perturbations.

Eq. (\ref{Ee2},\ref{Ee1}) become

\begin{equation}\label{EDisplacementXi}
-\boldsymbol{\zeta}\cdot\mathbf{C}^0:(\boldsymbol{\zeta}\otimes \mathbf{\dot{u}})_{sym}+\omega^2\rho^0\mathbf{\dot{u}}=\omega^2\rho^0\mathbf{\dot{U}}-\omega(\boldsymbol{\zeta}\cdot\boldsymbol{{\Sigma}})
\end{equation}

\begin{equation}\label{EStressXi}
-\mathbf{C}^0:[\boldsymbol{\zeta}\otimes(\boldsymbol{\zeta}\cdot\boldsymbol{\sigma})]_{sym}+\omega^2\rho^0\boldsymbol{\sigma}=\omega^2\rho^0\boldsymbol{\Sigma}-\omega\rho^0\mathbf{C}^0:(\boldsymbol{\zeta}\otimes\mathbf{\dot{U}})_{sym}
\end{equation}
where $\boldsymbol{\zeta}=\boldsymbol{\xi}+\mathbf{q}$. For the case of an isotropic reference material we have,

\begin{equation}
C^0_{ijkl}=\lambda^0\delta_{ij}\delta_{kl}+\mu^0[\delta_{ik}\delta_{jl}+\delta_{il}\delta_{jk}]
\end{equation}

Using the isotropic stiffness tensor, the tensors in Eqs. (\ref{EDisplacementXi}, \ref{EStressXi}) can be inverted to give, 

\begin{equation}\label{EInverted}
\begin{array}{l}
\displaystyle \mathbf{\dot{u}}(\boldsymbol{\xi})=\mathbf{\Phi}(\boldsymbol{\xi})\cdot\mathbf{\dot{U}}(\boldsymbol{\xi})+\mathbf{\Theta}(\boldsymbol{\xi}):\mathbf{\Sigma}(\boldsymbol{\xi})\\
\displaystyle \boldsymbol{{\sigma}}(\boldsymbol{\xi})=\mathbf{\Psi}(\boldsymbol{\xi})\cdot\mathbf{\dot{U}}(\boldsymbol{\xi})+\mathbf{\Gamma}(\boldsymbol{\xi}):\mathbf{\Sigma}(\boldsymbol{\xi})\\

\end{array} 
\end{equation}
where (see Appendix),

\begin{equation}
\Phi_{ij}=\omega^2\left[\frac{c_1^2-c_2^2}{\left[\omega^2-c_2^2{\zeta}^2\right]\left[\omega^2-c_1^2{\zeta}^2\right]}\zeta_i\zeta_j+\frac{1}{\left[\omega^2-c_2^2{\zeta}^2\right]}\delta_{ij}\right]
\end{equation}

\begin{equation}
\Theta_{ijp}=-\frac{1}{\omega\rho^0}\Phi_{ij}\zeta_p
\end{equation}

\begin{eqnarray}
  \Gamma_{ijkl} & = & \frac{c_2^2}{\omega^2-c_2^2{\zeta}^2}\frac{1}{2}\{\zeta_i\delta_{jk}\zeta_l+\zeta_i\delta_{jl}\zeta_k+\zeta_j\delta_{ik}\zeta_l+\zeta_j\delta_{il}\zeta_k\} \\
  & & {} +\frac{c_1^2-2c_2^2}{\omega^2-c_1^2\boldsymbol{\zeta}^2}\delta_{ij}\zeta_k\zeta_l\nonumber+\frac{2c_2^2[c_1^2-c_2^2]}{[\omega^2-c_2^2{\zeta}^2][\omega^2-c_1^2{\zeta}^2]}\zeta_i\zeta_j\zeta_k\zeta_l+\frac{1}{2}\left[\delta_{ik}\delta_{jl}+\delta_{il}\delta_{jk}\right]\nonumber
\end{eqnarray}

\begin{eqnarray}
  \Psi_{ijp} & = & -\omega\rho^0\left[\left\{\frac{2c_2^2(c_1^2-c_2^2)}{[\omega^2-c_2^2{\zeta}^2][\omega^2-c_1^2{\zeta}^2]}\right\}\zeta_i\zeta_j\zeta_p\right. \\
  & & {} \left.+\left\{\frac{c_1^2-2c_2^2}{\omega^2-c_1^2{\zeta}^2}\right\}\delta_{ij}\zeta_p+\left\{\frac{c_2^2}{\omega^2-c_2^2{\zeta}^2}\right\}\{\zeta_i\delta_{jp}+\zeta_j\delta_{ip}\}\right]\nonumber
\end{eqnarray}
where $c_1=\sqrt{(\lambda^0+2\mu^0)/\rho^0}$ is the longitudinal wave speed and $c_1=\sqrt{\mu^0/\rho^0}$ is the shear wave speed. It can be seen that $\boldsymbol{\Gamma}$ is symmetric about its first two and last two indices, $\boldsymbol{\Phi}$ is symmetric about its indices and $\boldsymbol{\Psi}$ is symmetric about its first two indices. $\boldsymbol{\Theta}$ is not automatically symmetric about its last two indices but since it contracts with eigenstress (Eq. \ref{EInverted}) which is a symmetric tensor, it can be symmetrized about its last two indices (the antisymmetric part vanishes when contracted with the symmetric eigenstress tensor),

\begin{equation}
{\Theta}_{ijp}=\frac{1}{2}(\Theta_{ijp}+\Theta_{ipj})
\end{equation}

Now the stress and velocity fields can be expressed as a sum of their average and their zero mean periodic components,

\begin{equation}\label{EStressPer}
\mathbf{\dot{u}}(\mathbf{x})=\langle \mathbf{\dot{u}}\rangle+\sum_{\boldsymbol{\xi}\neq 0}e^{i\boldsymbol{\xi}\cdot \mathbf{x}}\left[\mathbf{\Phi}(\boldsymbol{\xi})\cdot\frac{1}{\Omega}\int_{\Omega}\mathbf{\dot{U}}(\mathbf{y})e^{-i\boldsymbol{\xi}\cdot \mathbf{y}}dV_\mathbf{y}+\mathbf{\Theta}(\boldsymbol{\xi}):\frac{1}{\Omega}\int_{\Omega}\boldsymbol{\Sigma}(\mathbf{y})e^{-i\boldsymbol{\xi}\cdot \mathbf{y}}dV_\mathbf{y}\right]
\end{equation}

\begin{equation}\label{EDispPer}
\boldsymbol{\sigma}(\mathbf{x})=\langle \boldsymbol{\sigma}\rangle+\sum_{\boldsymbol{\xi}\neq 0}e^{i\boldsymbol{\xi}\cdot \mathbf{x}}\left[\mathbf{\Psi}(\boldsymbol{\xi})\cdot\frac{1}{\Omega}\int_{\Omega}\mathbf{\dot{U}}(\mathbf{y})e^{-i\boldsymbol{\xi}\cdot \mathbf{y}}dV_\mathbf{y}+\mathbf{\Gamma}(\boldsymbol{\xi}):\frac{1}{\Omega}\int_{\Omega}\boldsymbol{\Sigma}(\mathbf{y})e^{-i\boldsymbol{\xi}\cdot \mathbf{y}}dV_\mathbf{y}\right]
\end{equation}
where $\langle \mathbf{\dot{u}}\rangle$ and $\langle \boldsymbol{\sigma}\rangle$ are the average values of the velocity and stress fields, respectively, taken over a unit cell. 

To make the homogenized unit cell point-wise equivalent to the original heterogeneous unit cell, the homogenizing fields are required to satisfy the following consistency conditions:

\begin{equation}
\mathbf{D(x)}:[\langle\boldsymbol{\sigma}\rangle+\boldsymbol{\sigma}^p]=\mathbf{D}^0:[\langle\boldsymbol{\sigma}\rangle+\boldsymbol{\sigma}^p-\boldsymbol{\Sigma}]
\end{equation}

\begin{equation}
\rho(\mathbf{x})[\langle \mathbf{\dot{u}}\rangle+\mathbf{\dot{u}}^p]=\rho^0[\langle \mathbf{\dot{u}}\rangle+\mathbf{\dot{u}}^p-\mathbf{\dot{U}}]
\end{equation}
where $\mathbf{D}$ and $\mathbf{D}^0$ are the compliance tensors of the actual and the reference materials respectively. The periodic parts of the velocity and stress fields, from Eqs. (\ref{EDispPer}, \ref{EStressPer}), are now substituted into the above equations. 
This gives a set of 2 coupled integral equations which yields the required homogenizing stress and velocity fields that exactly and fully replace the heterogeneity in the original medium. 
Our immediate concern, however, is not the point-wise representation of the heterogeneous medium, but, rather, it is the determination of the averaged field values, though the solution technique also yields the point-wise values of the field values as well. 

 We now divide the unit cell into $\bar{\alpha}$ number of subregions, $\Omega_\alpha$. 
 Then, we average the perturbation fields over each such subregion to obtain,

\begin{eqnarray}\label{EStressXiAv}
\lefteqn{\langle \boldsymbol{\sigma}^p\rangle_{\Omega_\alpha}=\boldsymbol{\sigma}^{p\alpha}=\frac{1}{\Omega_\alpha}\int_{\Omega_\alpha}\boldsymbol{\sigma}^p(\mathbf{x})dV_\mathbf{x}}\\
& & \nonumber =\sum_{\boldsymbol{\xi}\neq 0}g^\alpha(\boldsymbol{\xi})\left[\mathbf{\Psi}(\boldsymbol{\xi})\cdot\frac{1}{\Omega}\int_{\Omega}\mathbf{\dot{U}}(\mathbf{y})e^{-i\boldsymbol{\xi}\cdot \mathbf{y}}dV_\mathbf{y}+\mathbf{\Gamma}(\boldsymbol{\xi}):\frac{1}{\Omega}\int_{\Omega}\boldsymbol{\Sigma}(\mathbf{y})e^{-i\boldsymbol{\xi}\cdot \mathbf{y}}dV_\mathbf{y}\right]
\end{eqnarray}

\begin{eqnarray}\label{EDisplacementXiAv}
\lefteqn{\langle \mathbf{\dot{u}}^p\rangle_{\Omega_\alpha}=\mathbf{\dot{u}}^{p\alpha}=\frac{1}{\Omega_\alpha}\int_{\Omega_\alpha}\mathbf{\dot{u}}^p(\mathbf{x})dV_\mathbf{x}}\\
& & \nonumber = \sum_{\boldsymbol{\xi}\neq 0}g^\alpha(\boldsymbol{\xi})\left[\mathbf{\Phi}(\boldsymbol{\xi})\cdot\frac{1}{\Omega}\int_{\Omega}\mathbf{\dot{U}}(\mathbf{y})e^{-i\boldsymbol{\xi}\cdot \mathbf{y}}dV_\mathbf{y}+\mathbf{\Theta}(\boldsymbol{\xi}):\frac{1}{\Omega}\int_{\Omega}\boldsymbol{\Sigma}(\mathbf{y})e^{-i\boldsymbol{\xi}\cdot \mathbf{y}}dV_\mathbf{y}\right]
\end{eqnarray}

\begin{equation}
g^\alpha(\boldsymbol{\xi})=\frac{1}{\Omega_\alpha}\int_{\Omega_\alpha}\mathrm{e}^{i\boldsymbol{\xi}\cdot\mathbf{x}}dV_\mathbf{x}
\end{equation}

We now replace the integrals in Eqs. (\ref{EStressXiAv}, \ref{EDisplacementXiAv}) by their equivalent finite sums and set,

\begin{equation}
\begin{array}{l}
\displaystyle \frac{1}{\Omega}\int_\Omega \mathbf{F(y)}\mathrm{e}^{-i\boldsymbol{\xi}\cdot\mathbf{y}}dV_\mathbf{y}\approx \sum_\beta f^\beta g^\beta(-\boldsymbol{\xi})\mathbf{F}^\beta\\
\displaystyle f^\beta=\frac{\Omega_\beta}{\Omega}\\
\displaystyle \mathbf{F}^\beta=\langle \mathbf{F}\rangle_{\Omega_\beta}
\end{array} 
\end{equation}

Eqs. (\ref{EStressXiAv}, \ref{EDisplacementXiAv}) then yield the following expressions:

\begin{equation}\label{EStressXiAvbeta}
\boldsymbol{\sigma}^{p\alpha}=\boldsymbol{\Psi}^{\alpha\beta}\cdot\mathbf{\dot{U}}^\beta+\boldsymbol{\Gamma}^{\alpha\beta}:\boldsymbol{\Sigma}^\beta
\end{equation}

\begin{equation}\label{EDisplacementXiAvbeta}
\mathbf{\dot{u}}^{p\alpha}=\boldsymbol{\Phi}^{\alpha\beta}\cdot\mathbf{\dot{U}}^\beta+\boldsymbol{\Theta}^{\alpha\beta}:\boldsymbol{\Sigma}^\beta
\end{equation}
where the repeated index, $\beta$, is summed over the number of subregions, ${\beta} = 1, \dots, \bar{\alpha}$, and greek indices serve as labels for tensors rather than components of a particular tensor.  The coefficient tensors in the above equations are defined by,

\begin{equation}
\mathbf{M}^{\alpha\beta}=\sum_{\xi\neq 0}g^\alpha(\boldsymbol{\xi})f^\beta g^\beta(-\boldsymbol{\xi})\mathbf{M}(\boldsymbol{\xi})
\end{equation}

In these equations, $\beta$ is not summed. Averaging the consistency conditions over each subregion $\alpha$ and using Eqs. (\ref{EStressXiAvbeta}, \ref{EDisplacementXiAvbeta}), we have,

\begin{equation}\label{EFinal}
\begin{array}{l}
\displaystyle f^{\alpha}\langle\boldsymbol{{\sigma}}\rangle=-\tilde{\boldsymbol{\Gamma}}^{\alpha\beta}:\boldsymbol{\Sigma}^\beta-\bar{\boldsymbol{\Psi}}^{\alpha\beta}\cdot\mathbf{{\dot{U}}}^\beta\\

\displaystyle f^{\alpha}\langle \mathbf{{\dot{u}}}\rangle=-\bar{\boldsymbol{\Theta}}^{\alpha\beta}:\boldsymbol{\Sigma}^\beta- \tilde{\boldsymbol{\Phi}}^{\alpha\beta}\cdot\mathbf{{\dot{U}}}^\beta\\

\displaystyle \bar{\mathbf{M}}^{\alpha\beta}=f^\alpha\mathbf{M}^{\alpha\beta}\:;\quad\quad\alpha\:not\:summed\\

\displaystyle \tilde{\boldsymbol{\Gamma}}^{\alpha\beta}=\left[\bar{\boldsymbol{\Gamma}}^{\alpha\beta}+f^\alpha\delta^{\alpha\beta} (\mathbf{D}^\alpha-\mathbf{D}^0)^{-1}:\mathbf{D}^0\right]\\

\tilde{\boldsymbol{\Phi}}^{\alpha\beta}=\left[\bar{\boldsymbol{\Phi}}^{\alpha\beta}+\mathbf{1}^{(2)}\frac{f^\alpha\rho^0}{\rho^\alpha-\rho^0}\delta^{\alpha\beta}\right]
\end{array} 
\end{equation}

For each tensor in Eq. (\ref{EFinal}) we have,

\begin{equation}
\begin{array}{l}
\displaystyle \bar{\mathbf{M}}^{\alpha\beta}=\sum_{\xi>0}\left[f^\alpha g^\alpha(\boldsymbol{\xi})f^\beta g^\beta(-\boldsymbol{\xi})\mathbf{M}(\boldsymbol{\xi})+f^\alpha g^\alpha(-\boldsymbol{\xi})f^\beta g^\beta(\boldsymbol{\xi})\mathbf{M}(-\boldsymbol{\xi})\right]\\

\displaystyle \bar{\mathbf{M}}^{\beta\alpha}=\sum_{\xi>0}\left[f^\beta g^\beta(\boldsymbol{\xi})f^\alpha g^\alpha(-\boldsymbol{\xi})\mathbf{M}(\boldsymbol{\xi})+f^\beta g^\beta(-\boldsymbol{\xi})f^\alpha g^\alpha(\boldsymbol{\xi})\mathbf{M}(-\boldsymbol{\xi})\right]\\

\displaystyle \bar{\mathbf{M}}^{\alpha\beta}=\left[\bar{\mathbf{M}}^{\beta\alpha}\right]^*\\

\end{array} 
\end{equation}
where $*$ denotes conjugation. This property also holds for tensors $\tilde{\boldsymbol{\Gamma}}^{\alpha\beta}$ and $\tilde{\boldsymbol{\Phi}}^{\alpha\beta}$ since they are derived from $\bar{\boldsymbol{\Gamma}}^{\alpha\beta}$ and $\bar{\boldsymbol{\Phi}}^{\alpha\beta}$ by modifying only those tensors for which $\alpha=\beta$. We also have the following identity (see Appendix),

\begin{equation}\label{EPsiTheta}
\hat{\Psi}^{\alpha\beta}_{mnp}=D^0_{mnqr}\bar{\Psi}^{\alpha\beta}_{qrp}=\rho^0\bar{\Theta}^{\alpha\beta}_{pmn}=\hat{\Theta}^{\alpha\beta}_{pmn}
\end{equation}

The above identity suggests that it is expedient to express Eq. (\ref{EFinal}) in terms of the modified tensors, $\hat{\boldsymbol{\Psi}}$ and $\hat{\boldsymbol{\Theta}}$. This can be achieved by multiplying the matrix corresponding to the tensor $\hat{\boldsymbol{\Psi}}$ with the stiffness matrix and multiplying the matrix corresponding to the tensor $\hat{\boldsymbol{\Theta}}$ with a diagonal matrix consisting of $1/\rho^0$ on its diagonals. We also transform tensor ${\boldsymbol{\Gamma}}$ in a similar way in order to yield $\hat{\boldsymbol{\Gamma}}$ which in addition to having minor symmetries over its first two and last two indices, also has major symmetry over them (See Appendix),

\begin{equation}
  \hat{\Gamma}_{mnkl}=D^0_{mnij}\tilde{\Gamma}_{ijkl}
\end{equation}

and transforming $\boldsymbol{\Phi}$ to

\begin{equation}
\hat{\Phi}_{ij}=\rho^0\tilde{\Phi}_{ij}
\end{equation}

To facilitate inversion and solution, Eqs. (\ref{EFinal}) are now expressed in their equivalent matrix form,

\begin{equation}\label{EMatrix}
\begin{array}{l}
\displaystyle \left[\mathbf{f}^1\right]\{\langle\boldsymbol{\sigma}\rangle\}=-\left[\bar{\mathbf{W}}\right]^{-1}\left[\bar{\mathbf{D}}^0\right]^{-1}\left[\hat{\boldsymbol{\Gamma}}\right]\left[\bar{\mathbf{W}}\right]\{\boldsymbol{\Sigma}\}-\left[\bar{\mathbf{W}}\right]^{-1}\left[\bar{\mathbf{D}}^0\right]^{-1}\left[\hat{\boldsymbol{\Psi}}\right]\{\dot{\mathbf{U}}\}\\
\displaystyle \left[\mathbf{f}^2\right]\{\langle\dot{\mathbf{u}}\rangle\}=-\left[\rho^0\right]^{-1}\left[\hat{\boldsymbol{\Theta}}\right]\left[\bar{\mathbf{W}}\right]\{\boldsymbol{\Sigma}\}-\left[\rho^0\right]^{-1}\left[\hat{\boldsymbol{\Phi}}\right]\{\dot{\mathbf{U}}\}\\

\displaystyle \{\langle\boldsymbol{\sigma}\rangle\}_{6(\alpha-1)+1}^{6(\alpha-1)+6}=\{\langle{\sigma}_{11}\rangle\;\langle{\sigma}_{22}\rangle\;\langle{\sigma}_{33}\rangle\;\langle{\sigma}_{23}\rangle\;\langle{\sigma}_{31}\rangle\;\langle{\sigma}_{12}\rangle\}^\mathrm{T}\\

\displaystyle \{\langle\dot{\mathbf{u}}\rangle\}_{3(\alpha-1)+1}^{3(\alpha-1)+3}=\{\langle\dot{{u}}_1\rangle\;\langle\dot{{u}}_2\rangle\;\langle\dot{{u}}_3\rangle\}^\mathrm{T}\\

\displaystyle \{\boldsymbol{\Sigma}\}_{6(\alpha-1)+1}^{6(\alpha-1)+6}=\{{\Sigma}^\alpha_{11}\;{\Sigma}^\alpha_{22}\;{\Sigma}^\alpha_{33}\;{\Sigma}^\alpha_{23}\;{\Sigma}^\alpha_{31}\;{\Sigma}^\alpha_{12}\}^\mathrm{T}\\

\displaystyle \{\dot{\mathbf{U}}\}_{3(\alpha-1)+1}^{3(\alpha-1)+3}=\{\dot{{U}}^\alpha_{1}\;\dot{{U}}^\alpha_{2}\;\dot{{U}}^\alpha_{3}\}^\mathrm{T}\\

\displaystyle \mathbf{f}^1_{ij}=f^\alpha\delta_{ij};\quad \mathbf{f}^2_{kl}=f^\alpha\delta_{kl}\\

\displaystyle i=6(\alpha-1)+1:6(\alpha-1)+6\\
\displaystyle j=6(\alpha-1)+1:6(\alpha-1)+6\\

\displaystyle k=3(\alpha-1)+1:3(\alpha-1)+3\\

\displaystyle l=3(\alpha-1)+1:3(\alpha-1)+3\\

\displaystyle \alpha=1:\bar{\alpha}\\

\end{array} 
\end{equation}
where$\left[\bar{\mathbf{D}}^0\right]$ is a $6\bar{\alpha}\times 6\bar{\alpha}$ symmetric matrix with the $6\times 6$ compliance matrix $\left[\mathbf{D}^0\right]$ about its diagonal and $\left[\bar{\mathbf{W}}\right]$ is a $6\bar{\alpha}\times 6\bar{\alpha}$ symmetric matrix with the $6\times 6$ diagonal matrix $\left[\mathbf{W}\right]$=diag(1 1 1 2 2 2) about its diagonal. Similarly, $\left[\rho^0\right]^{-1}=\left[1/\rho^0\right]$ is a $3\bar{\alpha}\times 3\bar{\alpha}$ symmetric matrix with the reciprocal of the reference material's density, $\rho^0$, on its diagonal.

In the above equations $\left[\hat{\boldsymbol{\Gamma}}\right]$ is a $6\bar{\alpha}\times 6\bar{\alpha}$ square matrix and $\left[\hat{\boldsymbol{\Phi}}\right]$ is a $3\bar{\alpha}\times 3\bar{\alpha}$ square matrix. For any $\alpha$ and $\beta$ we have, 

\begin{equation}\label{EHermitian}
\begin{array}{l}
\displaystyle \left[\hat{\boldsymbol{\Gamma}}\right]_{ij}= \hat{{\Gamma}}^{\alpha\beta}_{pqrs}=\{\hat{{\Gamma}}^{\beta\alpha}_{pqrs}\}^*=\{\hat{{\Gamma}}^{\beta\alpha}_{rspq}\}^*=\left[\hat{\boldsymbol{\Gamma}}\right]_{ji}^*\\

\displaystyle \left[\hat{\boldsymbol{\Phi}}\right]_{ij}= \hat{{\Phi}}^{\alpha\beta}_{pq}=\{\hat{{\Phi}}^{\beta\alpha}_{pq}\}^*=\{\hat{{\Phi}}^{\beta\alpha}_{qp}\}^*=\left[\hat{\boldsymbol{\Phi}}\right]_{ji}^*\\

\end{array} 
\end{equation}
where * denotes conjugation. The above result is general since for $\alpha=\beta$ tensors $\hat{\boldsymbol{\Gamma}}^{\alpha\alpha}$ and $\hat{\boldsymbol{\Phi}}^{\alpha\alpha}$ are purely real and symmetric so that, combined with Eq. (\ref{EHermitian}), the corresponding square matrices are hermitian.

Using Eq. (\ref{EPsiTheta}) we also note the following identity for the matrices $\left[\hat{\boldsymbol{\Psi}}\right]$ and $\left[\hat{\boldsymbol{\Theta}}\right]$,

\begin{equation}\label{EPsiThetahat}
\left[\hat{\boldsymbol{\Psi}}\right]_{ij}=\hat{\Psi}^{\alpha\beta}_{mnp}=\hat{\Theta}^{\alpha\beta}_{pmn}=\{\hat{\Theta}^{\beta\alpha}_{pmn}\}^*=\left[\hat{\boldsymbol{\Theta}}\right]_{ji}^*
\end{equation}

More generally, denoting a hermitian transpose by \textdagger, we have the following identities for the matrices,

\begin{equation}\label{ETranspose}
\begin{array}{l}
\displaystyle \left[\hat{\boldsymbol{\Gamma}}\right]^\dagger=\left[\hat{\boldsymbol{\Gamma}}\right]\\
\displaystyle \left[\hat{\boldsymbol{\Phi}}\right]^\dagger=\left[\hat{\boldsymbol{\Phi}}\right]\\
\displaystyle \left[\hat{\boldsymbol{\Psi}}\right]^\dagger=\left[\hat{\boldsymbol{\Theta}}\right];\quad\left[\hat{\boldsymbol{\Theta}}\right]^\dagger=\left[\hat{\boldsymbol{\Psi}}\right]\\
\end{array} 
\end{equation}

To expedite notation we omit the square brackets denoting a matix in further calculations. Curly braces and angle brackets indicating vectors and averaged vectors resectively are retained. We express the solution to Eq. (\ref{EMatrix}) in the following form:

\begin{equation}\label{EInvertedSolution}
\begin{array}{l}
\displaystyle \{\boldsymbol{\Sigma}\}=\boldsymbol{\Delta}\langle\boldsymbol{{\sigma}}\rangle+\boldsymbol{\Lambda}\langle\mathbf{{\dot{u}}}\rangle\\
\displaystyle \{\mathbf{\dot{U}}\}=\boldsymbol{\Xi}\langle\boldsymbol{{\sigma}}\rangle+\boldsymbol{\Omega}\langle\mathbf{{\dot{u}}}\rangle\\
\end{array} 
\end{equation}
where,

\begin{equation}
\begin{array}{l}
\displaystyle 
\boldsymbol{\Delta}=\left[-\bar{\mathbf{W}}^{-1}(\mathbf{\bar{D}}^0)^{-1}\hat{\boldsymbol{\Gamma}}\bar{\mathbf{W}}+\bar{\mathbf{W}}^{-1}(\mathbf{\bar{D}}^0)^{-1}\boldsymbol{\hat{\Psi}}\boldsymbol{\hat{\Phi}}^{-1}\boldsymbol{\hat{\Theta}}\mathbf{W}\right]^{-1}\mathbf{f}^1\\
\displaystyle {\boldsymbol{\Lambda}}=-\left[-\bar{\mathbf{W}}^{-1}(\mathbf{\bar{D}}^0)^{-1}\hat{\boldsymbol{\Gamma}}\bar{\mathbf{W}}+\right.\\
\quad\quad\quad\quad\left.\bar{\mathbf{W}}^{-1}(\mathbf{\bar{D}}^0)^{-1}\boldsymbol{\hat{\Psi}}\boldsymbol{\hat{\Phi}}^{-1}\boldsymbol{\hat{\Theta}}\mathbf{W}\right]^{-1}\bar{\mathbf{W}}^{-1}(\bar{\mathbf{D}}^0)^{-1}\hat{\boldsymbol{\Psi}}\hat{\boldsymbol{\Phi}}^{-1}\bar{\boldsymbol{\rho}}^0\mathbf{f}^2\\

\displaystyle {\boldsymbol{\Xi}}=-\left[-(\bar{\boldsymbol{\rho}}^0)^{-1}\hat{\boldsymbol{\Phi}}+(\bar{\boldsymbol{\rho}}^0)^{-1}\hat{\boldsymbol{\Theta}}\hat{\boldsymbol{\Gamma}}^{-1}\hat{\boldsymbol{\Psi}}\right]^{-1}(\bar{\boldsymbol{\rho}}^0)^{-1}\hat{\boldsymbol{\Theta}}\hat{\boldsymbol{\Gamma}}^{-1}\bar{\mathbf{D}}^0\bar{\mathbf{W}}\mathbf{f}^1\\
\displaystyle
\boldsymbol{\Omega}=\left[-(\bar{\boldsymbol{\rho}}^0)^{-1}\hat{\boldsymbol{\Phi}}+(\bar{\boldsymbol{\rho}}^0)^{-1}\hat{\boldsymbol{\Theta}}\hat{\boldsymbol{\Gamma}}^{-1}\hat{\boldsymbol{\Psi}}\right]^{-1}\mathbf{f}^2\\

\end{array} 
\label{EFinalConstR}
\end{equation}

Eqs. (\ref{EInvertedSolution}) expresses the vectors of eigenstress and eigenvelocity in each subdivision in terms of the vector of average stress and average velocity. It can be seen from Eq. (\ref{EFinal}) that the vectors of the average quantities in Eqs. (\ref{EInvertedSolution}) are composed of $\bar{\alpha}$ times repeated vectors of the averaged quantities. Therefore, Eqs. (\ref{EInvertedSolution}) can be condensed to express the vectors of eigenstress and eigenvelocity in each subdivision in terms a $6\times 1$ vector of average stress and a $3\times 1$ vector of average velocity. Furthermore, we can also average the vectors of eigenstress and eigenvelocity over all the subregions to finally get,

\begin{equation}\label{EInvertedSolutionAv}
\begin{array}{l}
\displaystyle \{\boldsymbol{\langle\Sigma}\rangle\}_{6\times 1}=\langle\boldsymbol{\Delta}\rangle_{6\times 6}\langle\boldsymbol{{\sigma}}\rangle_{6\times 1}+\langle\boldsymbol{\Lambda}\rangle_{6\times 3}\langle\mathbf{{\dot{u}}}\rangle_{3\times 1}\\
\displaystyle \{\langle\mathbf{\dot{U}}\rangle\}_{3\times 1}=\langle\boldsymbol{\Xi}\rangle_{3\times 6}\langle\boldsymbol{{\sigma}}\rangle_{6\times 1}+\langle\boldsymbol{\Omega}\rangle_{3\times 3}\langle\mathbf{{\dot{u}}}\rangle_{3\times 1}\\
\end{array} 
\end{equation}
where,

\begin{equation}
\begin{array}{l}
\displaystyle 
\langle\boldsymbol{\Delta}\rangle=(\mathbf{F}^1)^T\left[-\bar{\mathbf{W}}^{-1}(\mathbf{\bar{D}}^0)^{-1}\hat{\boldsymbol{\Gamma}}\bar{\mathbf{W}}+\bar{\mathbf{W}}^{-1}(\mathbf{\bar{D}}^0)^{-1}\boldsymbol{\hat{\Psi}}\boldsymbol{\hat{\Phi}}^{-1}\boldsymbol{\hat{\Theta}}\bar{\mathbf{W}}\right]^{-1}\mathbf{F}^1\\
\displaystyle \langle{\boldsymbol{\Lambda}}\rangle=-(\mathbf{F}^1)^T\left[-\bar{\mathbf{W}}^{-1}(\mathbf{\bar{D}}^0)^{-1}\hat{\boldsymbol{\Gamma}}\bar{\mathbf{W}}+\right.\\
\quad\quad\quad\quad\left.\bar{\mathbf{W}}^{-1}(\mathbf{\bar{D}}^0)^{-1}\boldsymbol{\hat{\Psi}}\boldsymbol{\hat{\Phi}}^{-1}\boldsymbol{\hat{\Theta}}\bar{\mathbf{W}}\right]^{-1}\bar{\mathbf{W}}^{-1}(\bar{\mathbf{D}}^0)^{-1}\hat{\boldsymbol{\Psi}}\hat{\boldsymbol{\Phi}}^{-1}\bar{\boldsymbol{\rho}}^0\mathbf{F}^2\\

\displaystyle \langle{\boldsymbol{\Xi}}\rangle=-(\mathbf{F}^2)^T\left[-(\bar{\boldsymbol{\rho}}^0)^{-1}\hat{\boldsymbol{\Phi}}+(\bar{\boldsymbol{\rho}}^0)^{-1}\hat{\boldsymbol{\Theta}}\hat{\boldsymbol{\Gamma}}^{-1}\hat{\boldsymbol{\Psi}}\right]^{-1}(\bar{\boldsymbol{\rho}}^0)^{-1}\hat{\boldsymbol{\Theta}}\hat{\boldsymbol{\Gamma}}^{-1}\bar{\mathbf{D}}^0\bar{\mathbf{W}}\mathbf{F}^1\\
\displaystyle
\langle\boldsymbol{\Omega}\rangle=(\mathbf{F}^2)^T\left[-(\bar{\boldsymbol{\rho}}^0)^{-1}\hat{\boldsymbol{\Phi}}+(\bar{\boldsymbol{\rho}}^0)^{-1}\hat{\boldsymbol{\Theta}}\hat{\boldsymbol{\Gamma}}^{-1}\hat{\boldsymbol{\Psi}}\right]^{-1}\mathbf{F}^2\\

\end{array} 
\label{EAveragedMatrices}
\end{equation}
where,

\[ \left[\mathbf{F}^1\right] = \left[ \begin{array}{cccccc}
f^1 & 0 & 0 & 0 & 0 & 0 \\
0 & f^1 & 0 & 0 & 0 & 0\\
0 & 0 & f^1 & 0 & 0 & 0\\
. & . & . & . & . & .\\
f^2 & 0 & 0 & 0 & 0 & 0\\
. & . & . & . & . & . \end{array} \right]\]

\[ \left[\mathbf{F}^2\right] = \left[ \begin{array}{ccc}
f^1 & 0 & 0 \\
0 & f^1 & 0 \\
0 & 0 & f^1 \\
f^2 & 0 & 0 \\
. & . & . \end{array} \right]\]

Now we average the consistency conditions over the entire unit cell to express the average strain and average momentum in terms of the average stress and average velocity tensors. Noting that the average of the periodic parts vanish when taken over the entire unit cell, we have,

\begin{equation}\label{e15}
\langle\boldsymbol{\varepsilon}\rangle=\mathbf{D}^0\mathbf{W}\left[\langle\boldsymbol{\sigma}\rangle-\langle\boldsymbol{\Sigma}\rangle\right]=\mathbf{\bar{D}}\langle\boldsymbol{\sigma}\rangle+\mathbf{\bar{S}}_1\langle \mathbf{\dot{u}}\rangle
\end{equation}

\begin{equation}\label{e16}
\langle \mathbf{p}\rangle=\rho^0[\langle \mathbf{\dot{u}}\rangle-\langle{\mathbf{\dot{U}}}\rangle]=\mathbf{\bar{S}}_2\langle\boldsymbol{\sigma}\rangle+\boldsymbol{\bar{\rho}}\langle\mathbf{\dot{u}}\rangle
\end{equation}

The effective parameters are given by,

\begin{equation}
\begin{array}{l}
\displaystyle \mathbf{\bar{D}}=\mathbf{D}^0\mathbf{W}\left[\mathbf{1}-\langle\boldsymbol{\Delta}\rangle\right]\\
\displaystyle \mathbf{\bar{S}}_1=-\mathbf{D}^0\mathbf{W}\langle\boldsymbol{\Lambda}\rangle\\
\displaystyle \mathbf{\bar{S}}_2\mathbf{W}=-\rho^0\langle\boldsymbol{\Xi}\rangle\\
\displaystyle \boldsymbol{\bar{\rho}}=\rho_0[\mathbf{1}-\langle\boldsymbol{\Omega}\rangle]\\
\end{array} 
\label{EEffective}
\end{equation}

Eqs. (\ref{e15}, \ref{e16}) are our final constitutive relations for the homogenized composite. These equations can be expressed in the following tensorial form,

\begin{equation}\label{e151}
\langle\boldsymbol{\varepsilon}\rangle=\mathbf{\bar{D}}:\langle\boldsymbol{\sigma}\rangle+\mathbf{\bar{S}}_1\cdot\langle \mathbf{\dot{u}}\rangle
\end{equation}

\begin{equation}\label{e161}
\langle \mathbf{p}\rangle=\mathbf{\bar{S}}_2:\langle\boldsymbol{\sigma}\rangle+\boldsymbol{\bar{\rho}}\cdot\langle\mathbf{\dot{u}}\rangle
\end{equation}

It is shown in the following section that the effective compliance tensor $\mathbf{\bar{D}}$ is such that $\bar{D}_{ijkl}=\bar{D}_{klij}^*$ and the effective density tensor $\bar{\boldsymbol{\rho}}$ is such that $\bar{\rho}_{ij}=\bar{\rho}_{ji}^*$. It is also shown that the coupling tensors have the relation $(\bar{S}_2)_{ijk}^*=(\bar{S}_1)_{jki}$. The effective constitutive relation can alternatively be expressed as (See Willis 2009, 2011),

\begin{equation}\label{e152}
\langle\boldsymbol{\sigma}\rangle=\mathbf{\bar{C}}:\langle\boldsymbol{\varepsilon}\rangle+\mathbf{{S}}\cdot\langle \mathbf{\dot{u}}\rangle
\end{equation}

\begin{equation}\label{e162}
\langle \mathbf{p}\rangle=\mathbf{\bar{S}}:\langle\boldsymbol{\varepsilon}\rangle+\boldsymbol{\bar{\rho}}^1\cdot\langle\mathbf{\dot{u}}\rangle
\end{equation}
where $\mathbf{\bar{C}}=\mathbf{\bar{D}}^{-1}$, $\mathbf{S}=-\mathbf{\bar{C}}:\mathbf{\bar{S}}_1$, $\mathbf{\bar{S}}=\mathbf{\bar{S}}_2:\mathbf{\bar{C}}$, and $\boldsymbol{\bar{\rho}}^1=\boldsymbol{\bar{\rho}}-\mathbf{\bar{S}}_2:\mathbf{\bar{C}}:\mathbf{\bar{S}}_1$. Willis has shown that the constitutive relation expressed in the above form has a self-adjoint structure, i.e., $\mathbf{\bar{S}}(\mathbf{q},\omega)=\mathbf{{S}}(-\mathbf{q},\omega)$. Using the properties of the effective parameters proved in the next section, it can be shown that the self-adjointness of the structure of the constitutive relations of Eqs. (\ref{e161},\ref{e162}) emerges identically from our formulation. 

\subsection{Properties of the Constitutive Parameters}

We note that the $6\times 6$ matrix $\mathbf{D}^0$ in Eq. (\ref{EEffective}) can be taken inside the matrix multiplication in Eq. (\ref{EAveragedMatrices}) by converting it to a $6\bar{\alpha}\times 6\bar{\alpha}$ matrix $\bar{\mathbf{D}}^0$ which is composed of $\mathbf{D}^0$ about its diagonal. Similarly $\mathbf{W}$ is taken inside the matrices by converting it to $\bar{\mathbf{W}}$ and $\rho^0$ is taken inside the matrices by converting it to $\bar{\boldsymbol{\rho}}^0$.

We have,

\begin{eqnarray}
\mathbf{\bar{D}} & = & \mathbf{D}^0\mathbf{W}\left[\mathbf{1}-\langle\boldsymbol{\Delta}\rangle\right] \nonumber \\
   & = & \mathbf{D}^0\mathbf{W}-\mathbf{D}^0\mathbf{W}\langle\boldsymbol{\Delta}\rangle\nonumber \\
   & = & \mathbf{D}^0\mathbf{W}-\mathbf{D}^0\mathbf{W}(\mathbf{F}^1)^T\left[-\bar{\mathbf{W}}^{-1}(\mathbf{\bar{D}}^0)^{-1}\hat{\boldsymbol{\Gamma}}\bar{\mathbf{W}}+\bar{\mathbf{W}}^{-1}(\mathbf{\bar{D}}^0)^{-1}\boldsymbol{\hat{\Psi}}\boldsymbol{\hat{\Phi}}^{-1}\boldsymbol{\hat{\Theta}}\bar{\mathbf{W}}\right]^{-1}\mathbf{F}^1 \nonumber \\
   & = & \mathbf{D}^0\mathbf{W}-(\mathbf{F}^1)^T\left[-\bar{\mathbf{W}}^{-1}(\mathbf{\bar{D}}^0)^{-1}\hat{\boldsymbol{\Gamma}}(\mathbf{\bar{D}}^0)^{-1}+\bar{\mathbf{W}}^{-1}(\mathbf{\bar{D}}^0)^{-1}\boldsymbol{\hat{\Psi}}\boldsymbol{\hat{\Phi}}^{-1}\boldsymbol{\hat{\Theta}}(\mathbf{\bar{D}}^0)^{-1}\right]^{-1}\mathbf{F}^1\nonumber \\
   & = & \mathbf{D}^0\mathbf{W}-(\mathbf{F}^1)^T\left[-(\mathbf{\bar{D}}^0)^{-1}\hat{\boldsymbol{\Gamma}}(\mathbf{\bar{D}}^0)^{-1}+(\mathbf{\bar{D}}^0)^{-1}\boldsymbol{\hat{\Psi}}\boldsymbol{\hat{\Phi}}^{-1}\boldsymbol{\hat{\Theta}}(\mathbf{\bar{D}}^0)^{-1}\right]^{-1}\mathbf{F}^1\mathbf{W}\nonumber \nonumber \\
\end{eqnarray}

Considering Eqs. (\ref{ETranspose}) and the fact that $\mathbf{D}^0$ and $\mathbf{W}$ are hermitian, the above treatment shows that $\mathbf{\bar{D}}$ is also hermitian. It can be shown that for a hermitian matrix $\mathbf{M}$, transformations of the form $(\mathbf{F}^1)^T\mathbf{M}\mathbf{F}^1$ or $(\mathbf{F}^2)^T\mathbf{M}\mathbf{F}^2$ result in hermitian matrices. This combined with the fact that $\mathbf{D}^0\mathbf{W}$ is real and symmetric proves that $\mathbf{\bar{D}}$ is also hermitian. Similarly it can be shown that $\boldsymbol{\bar{\rho}}$ is also a hermitian matrix.

Taking a hermitian transpose of Eq. (\ref{EEffective})$^3$ and multiplying both sides by $\mathbf{W}^{-1}$ on the left we have,

\begin{eqnarray}
  \mathbf{\bar{S}}_2^\dagger & = & -\mathbf{W}^{-1}\langle\boldsymbol{\Xi}\rangle^{\dagger}\rho^0 \nonumber \\
   & = & \mathbf{W}^{-1}\left[(\mathbf{F}^2)^T\left[-(\bar{\boldsymbol{\rho}}^0)^{-1}\hat{\boldsymbol{\Phi}}+(\bar{\boldsymbol{\rho}}^0)^{-1}\hat{\boldsymbol{\Theta}}\hat{\boldsymbol{\Gamma}}^{-1}\hat{\boldsymbol{\Psi}}\right]^{-1}(\bar{\boldsymbol{\rho}}^0)^{-1}\hat{\boldsymbol{\Theta}}\hat{\boldsymbol{\Gamma}}^{-1}\bar{\mathbf{D}}^0\bar{\mathbf{W}}\mathbf{F}^1\right]^{\dagger}\rho^0 \nonumber \\
   & = & \left[(\mathbf{F}^2)^T\bar{\boldsymbol{\rho}}^0\left[-(\bar{\boldsymbol{\rho}}^0)^{-1}\hat{\boldsymbol{\Phi}}+(\bar{\boldsymbol{\rho}}^0)^{-1}\hat{\boldsymbol{\Theta}}\hat{\boldsymbol{\Gamma}}^{-1}\hat{\boldsymbol{\Psi}}\right]^{-1}(\bar{\boldsymbol{\rho}}^0)^{-1}\hat{\boldsymbol{\Theta}}\hat{\boldsymbol{\Gamma}}^{-1}\bar{\mathbf{D}}^0\mathbf{F}^1\right]^{\dagger} \nonumber \\
   & = & \left[(\mathbf{F}^2)^T\bar{\boldsymbol{\rho}}^0\left[-\hat{\boldsymbol{\Gamma}}\hat{\boldsymbol{\Theta}}^{-1}\hat{\boldsymbol{\Phi}}+\hat{\boldsymbol{\Psi}}\right]^{-1}\bar{\mathbf{D}}^0\mathbf{F}^1\right]^{\dagger}\nonumber \\
   & = & (\mathbf{F}^1)^T\bar{\mathbf{D}}^0\left[-\hat{\boldsymbol{\Phi}}\hat{\boldsymbol{\Psi}}^{-1}\hat{\boldsymbol{\Gamma}}+\hat{\boldsymbol{\Theta}}\right]^{-1}\bar{\boldsymbol{\rho}}^0\mathbf{F}^2\nonumber \nonumber
\end{eqnarray}
and,

\begin{eqnarray}
\mathbf{\bar{S}}_1 & = & -\mathbf{D}^0\mathbf{W}\langle\boldsymbol{\Lambda}\rangle
\nonumber \\
   & = & 
\mathbf{D}^0\mathbf{W}(\mathbf{F}^1)^T\left[-\bar{\mathbf{W}}^{-1}(\mathbf{\bar{D}}^0)^{-1}\hat{\boldsymbol{\Gamma}}\bar{\mathbf{W}}\right. 
\nonumber \\
   &   &  \left.\quad\quad+\bar{\mathbf{W}}^{-1}(\mathbf{\bar{D}}^0)^{-1}\boldsymbol{\hat{\Psi}}\boldsymbol{\hat{\Phi}}^{-1}\boldsymbol{\hat{\Theta}}\bar{\mathbf{W}}\right]^{-1}\bar{\mathbf{W}}^{-1}(\bar{\mathbf{D}}^0)^{-1}\hat{\boldsymbol{\Psi}}\hat{\boldsymbol{\Phi}}^{-1}\bar{\boldsymbol{\rho}}^0\mathbf{F}^2
\nonumber \\
   & = & 
(\mathbf{F}^1)^T\bar{\mathbf{D}}^0\bar{\mathbf{W}}\left[-\hat{\boldsymbol{\Phi}}\hat{\boldsymbol{\Psi}}^{-1}\hat{\boldsymbol{\Gamma}}\bar{\mathbf{W}}+\hat{\boldsymbol{\Theta}}\mathbf{W}\right]^{-1}\bar{\boldsymbol{\rho}}^0\mathbf{F}^2
\nonumber \\
   & = &  (\mathbf{F}^1)^T\bar{\mathbf{D}}^0\left[-\hat{\boldsymbol{\Phi}}\hat{\boldsymbol{\Psi}}^{-1}\hat{\boldsymbol{\Gamma}}+\hat{\boldsymbol{\Theta}}\right]^{-1}\bar{\boldsymbol{\rho}}^0\mathbf{F}^2
\nonumber \\
\end{eqnarray}
proving that $\mathbf{\bar{S}}_2^\dagger=\mathbf{\bar{S}}_1$. Since the matrices $\left[\bar{\mathbf{D}}\right]$ and $\left[\bar{\boldsymbol{\rho}}\right]$ are hermitian, we have $\bar{D}_{ijkl}=\bar{D}_{klij}^*$ and  $\bar{\rho}_{ij}=\bar{\rho}_{ji}^*$ where * denotes a complex conjugate. This means that the scalar given by $a=\langle\sigma\rangle_{ij}\bar{D}_{ijkl}^*\langle\sigma\rangle_{kl}^*=\langle\sigma\rangle_{ij}\bar{D}_{klij}\langle\sigma\rangle_{kl}^*$ has a complex conjugate $a^*=\langle\sigma\rangle_{kl}\bar{D}_{klij}^*\langle\sigma\rangle_{ij}^*$. Since the pairs $i,j$ and $k,l$ are symbols upon which summation is carried out, we find that $a=a^*$ or that $a$ is real. Similarly $\langle\dot{u}\rangle_{i}\bar{\rho}_{ij}^*\langle\dot{u}\rangle_{j}^*$ can be shown to be a real scalar. The relation $\mathbf{\bar{S}}_2^\dagger=\mathbf{\bar{S}}_1$ implies that $(\bar{S}_2)_{ijk}^*=(\bar{S}_1)_{jki}$. Now energy which can be written as,

\begin{eqnarray}
  E & = &  \frac{1}{2}\left[\langle\sigma\rangle_{ij}\langle\varepsilon\rangle_{ij}^*+\langle\dot{u}\rangle_{i}\langle p\rangle_{i}^*\right]\\
   & = & \frac{1}{2}\left[\langle\sigma\rangle_{ij}\bar{D}_{ijkl}^*\langle\sigma\rangle_{kl}^* + \langle\sigma\rangle_{ij}(\bar{S}_1)_{ijk}^*\langle\dot{u}\rangle_{k}^* +\langle\dot{u}\rangle_{i}(\bar{S}_2)_{ijk}^*\langle\sigma\rangle_{jk}^* + \langle\dot{u}\rangle_{i}\bar{\rho}_{ij}^*\langle\dot{u}\rangle_{j}^*\right]
\nonumber \\
   & = & \frac{1}{2}\left[\langle\sigma\rangle_{ij}\bar{D}_{ijkl}^*\langle\sigma\rangle_{kl}^* + \langle\sigma\rangle_{ij}(\bar{S}_1)_{ijk}^*\langle\dot{u}\rangle_{k}^* +\langle\dot{u}\rangle_{i}(\bar{S}_1)_{jki}\langle\sigma\rangle_{jk}^* + \langle\dot{u}\rangle_{i}\bar{\rho}_{ij}^*\langle\dot{u}\rangle_{j}^*\right]
\nonumber \\
   & = & \frac{1}{2}\left[\langle\sigma\rangle_{ij}\bar{D}_{ijkl}^*\langle\sigma\rangle_{kl}^* + \left[\langle\sigma\rangle_{ij}(\bar{S}_1)_{ijk}^*\langle\dot{u}\rangle_{k}^* + (\langle\sigma\rangle_{ij}(\bar{S}_1)_{ijk}^*\langle\dot{u}\rangle_{k}^*)^*\right] + \langle\dot{u}\rangle_{i}\bar{\rho}_{ij}^*\langle\dot{u}\rangle_{j}^*\right]
\nonumber \\
\end{eqnarray}
is proved to be real-valued. Taking into account the symmetries of the homogenized constitutive parameters, it must be noted that the averaged constitutive relations of Eqs. (\ref{e15}, \ref{e16}) can have 45 independent constants at the maximum. Depending upon the material properties of the constituents and the symmetry of the unit cell, the number of independent constants will vary. The essential relations of the constitutive tensors proved in this section will hold regardless of the material anisotropy or directional asymmetry.

\section{Numerical Example: Elastic cubes periodically distributed in an elastic matrix}

\begin{figure}[htp]
\centering
\includegraphics[scale=.6]{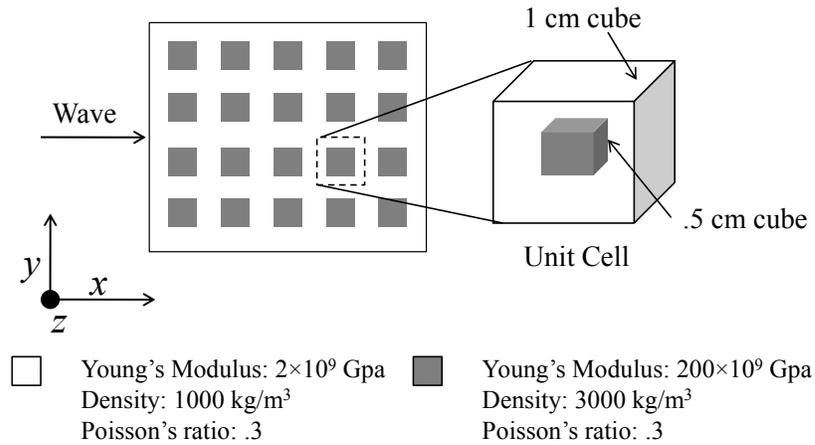}
\caption{Schematic of a 3-D periodic composite}\label{FSchematic}
\end{figure}

We consider a composite with a 3-D periodic microstructure composed of the periodically distributed unit cell shown in Fig. (\ref{FSchematic}). 

The composite is periodic in all three directions and the wave is assumed to propagate in the positive $x$ direction. Due to the periodicity of the composite, Bloch waves with wavenumber $q$ in the $x$ direction propagate in the structure. The wavenumber is related to the frequency by the dispersion relation. 

\begin{figure}[htp]
\centering
\includegraphics[scale=.6]{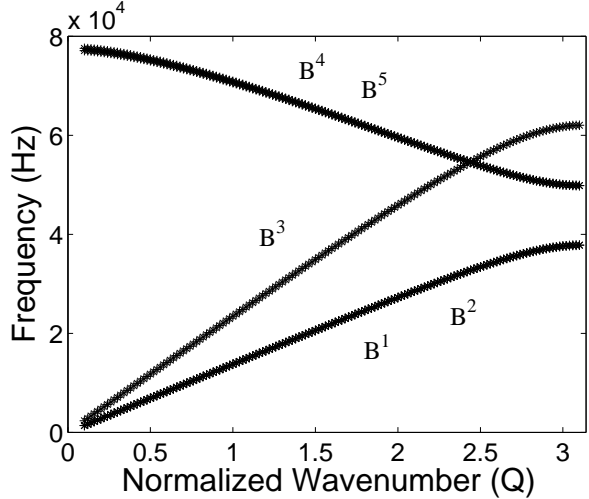}
\caption{Dispersion curve showing the first five modes of the 3-D composite}\label{FDispersion}
\end{figure}

Fig. (\ref{FDispersion}) shows the dispersion results for the first 5 modes calculated from the mixed variational formulation (See Iwakuma \& Nemat-Nasser 1982, Nemat-Nasser \& Srivastava 2011, and Nemat-Nasser \emph{et. al.} 2011). $B^1$ and $B^2$  indicate positive shear modes and $B^3$ indicates positive  longitudinal mode. $B^4$ and $B^5$ are shear modes for which the group velocity is anti-parallel to the phase velocity and are called negative shear modes.
$B^1$ and $B^2$ ($B^4$ and $B^5$) lie on top of each other because due to the symmetry of the unit cell, the shear directions, $y$ and $z$, are indistinguishable for a wave propagating in the $x$ direction.

Table (\ref{tab1}) shows the effective properties calculated from the micromechanical homogenization method proposed above. The effective properties are calculated for $Q=1.5$ on the positive longitudinal branch of the dispersion curve ($B^3$). The results were calculated by taking the reference material to be the matrix so that discretization of the inclusion was sufficient under the current method. The inclusion was discretized into 27 smaller cubes ($\bar{\alpha}=27$) within which the eigenstresses and eigenvelocities were approximated as constants. A total of $23^3-1$ fourier terms were used in the expansion of the periodic parts of the field variables.

It should be noted that all the components of the 4 effective tensors are real. This is due to the fact that the unit cell is symmetric in all three orthogonal directions. The essential relation between the coupling tensors ($(\bar{\mathbf{S}}^1)^\dagger=\bar{\mathbf{S}}^2$) is verified from the present calculation. The effective compliance tensor $\bar{\mathbf{D}}$ has both major and minor symmetries ($\bar{D}_{ijkl}=\bar{D}_{ijlk}=\bar{D}_{jikl}=\bar{D}_{klij}$) due to the symmetry of the problem. In general we should have $\bar{D}_{ijkl}=\bar{D}_{klij}^*$. Due to the symmetry of the unit cell, the effective density matrix is diagonal and indicates that there is no coupling between the average momentum in a particular direction and the average velocity in an orthogonal direction. The components of all the four effective tensors indicate that there is a distinction between the $x$ direction and the other two directions by virtue of the fact that the propagation direction is $x$. On the other hand the other two directions, $y$ and $z$, are indistinguishable for a wave propagating in the $x$ direction due to the symmetry of the unit cell. This is seen to be true from the calculated components of the effective tensors (eg. $\bar{D}_{2222}=\bar{D}_{3333}\neq\bar{D}_{1111}$ etc.)

\begin{table}\label{tab1}
\begin{center}
\begin{tabular}{ |c| c| c| c| c| c| c| c| c|}
\hline 
\multicolumn{6}{|c|}{$\bar{\mathbf{D}} (\times 10^{-10}Pa^{-1})$} & \multicolumn{3}{|c|}{$\bar{\mathbf{S}}^1 (\times 10^{-6}s/m)$} \\ 
\hline
3.8 & -1.08 & -1.08 & 0 & 0 & 0 & 6.76 & 0 & 0 \\ 
\hline
-1.08 & 3.75 & -1.05 & 0 & 0 & 0 & -3.47 & 0 & 0 \\ 
\hline
-1.08 & -1.05 & 3.75 & 0 & 0 & 0 & -3.47 & 0 & 0 \\ 
\hline
0 & 0 & 0 & 5.19 & 0 & 0 & 0 & 0 & 0 \\ 
\hline
0 & 0 & 0 & 0 & 5.2 & 0 & 0 & 0 & -.49 \\ 
\hline
0 & 0 & 0 & 0 & 0 & 5.2 & 0 & -.49 & 0 \\ 
\hline
\multicolumn{6}{|c|}{$\bar{\mathbf{S}}^2 (\times 10^{-6}m^2s)$} & \multicolumn{3}{|c|}{$\bar{\boldsymbol{\rho}} (kg/m^3)$} \\ 
\hline
6.76 & -3.47 & -3.47 & 0 & 0 & 0 & 393.5 & 0 & 0 \\ 
\hline
0 & 0 & 0 & 0 & 0 & -.49 & 0 & 395.5 & 0 \\ 
\hline
0 & 0 & 0 & 0 & -.49 & 0 & 0 & 0 & 395.5 \\ 
\hline
\end{tabular}\caption{Effective properties calculated for $Q=1.5$ on branch $B^3$}
\end{center}
\end{table}

\section{Conclusions}
A method for the homogenization of 3-D periodic elastic composites is presented. It allows for the evaluation of the overall dynamic material constitutive tensors relating the averaged dynamic field variable tensors of velocity, strain, stress, and linear momentum. The method doesn't require the apriori calculation of the solution to the elastodynamic problem. Rather, the dispersion relation and the pointwise solution of the composite emerge from the current method. The coupled form of the constitutive relation as proposed by Milton and Willis emerge naturally from the present method. We have also shown that the matrices corresponding to the effective comliance and density tensors are hermitian and that the coupling tensors are hermitian transose of each other. Finally we have presented corroboration of the theoretical proofs in the form of a solved example of a 3-D periodic composite composed of elastic cubes periodically distributed in an elastic matrix.

\section{Acknowledgement}

The authors are grateful to Professor John R. Willis for valuable discussions.  In particular, private communications with Dr. Willis served to establish the equivalence of the constitutive relations as presented in this paper and the constitutive relations as predicted by Dr. Willis' ensemble average method. This research has been conducted at the Center of Excellence for Advanced Materials (CEAM) at the University of California, San Diego, under DARPA, AFOSR Grant FA9550-09-1-0709.

\newpage
\appendix\label{Appendix}
\section{Appendix}
For isotropic case,

\begin{equation}
C^0_{ijkl}=\lambda^0\delta_{ij}\delta_{kl}+\mu^0[\delta_{ik}\delta_{jl}+\delta_{il}\delta_{jk}]
\end{equation}

The left hand side of Eq. (\ref{EDisplacementXi}) can be expanded as,

\begin{eqnarray}\label{EFFu}
 \lefteqn{-\boldsymbol{\zeta}\cdot\mathbf{C}^0:(\boldsymbol{\zeta}\otimes \mathbf{\dot{u}})_{sym}+\omega^2\rho^0\mathbf{\dot{u}} = H_{ij}\dot{u}_{j}=}\\
   && \nonumber\left[-(\lambda^0+\mu^0)\zeta_i\zeta_j+(\omega^2\rho^0-\mu^0\boldsymbol{\zeta}^2)\delta_{ij}\right]\dot{u}_{j}
\end{eqnarray}

To invert the tensor $\mathbf{H}$ in Eq. (\ref{EFFu}) we seek a tensor \textbf{J} such that,

\begin{equation}
\mathbf{J}_{mi}:\mathbf{H}_{ij}=\mathbf{1}^{2}_{mj}=\delta_{mj}
\end{equation}

Expressing $\mathbf{J}$ in terms of $\boldsymbol{\zeta}\otimes\boldsymbol{\zeta}$ and $\mathbf{1}^{2}$ we have,

\begin{equation}
\mathbf{J}=a(\boldsymbol{\zeta}\otimes\boldsymbol{\zeta})+b\mathbf{1}^{2}
\end{equation}

Contracting with $\mathbf{H}$ we get the following equation,

\begin{equation}\nonumber
\left[-a(\lambda^0+\mu^0)\boldsymbol{\zeta}^2+a(\omega^2\rho^0-\mu^0\boldsymbol{\zeta}^2)-b(\lambda^0+\mu^0)\right]\zeta_m\zeta_j+b(\omega^2\rho^0-\mu^0\boldsymbol{\zeta}^2)\delta_{mj}=\delta_{mj}
\end{equation}

The constants, therefore, are,

\begin{equation}
\begin{array}{l}
\displaystyle a=\frac{\lambda^0+\mu^0}{\left[\omega^2\rho^0-\mu^0\boldsymbol{\zeta}^2\right]\left[\omega^2\rho^0-(\lambda^0+2\mu^0)\boldsymbol{\zeta}^2\right]}\\
\displaystyle b=\frac{1}{\omega^2\rho^0-\mu^0\boldsymbol{\zeta}^2}\\
\end{array} 
\end{equation}

Now Eq. (\ref{EDisplacementXi}) can be written as,

\begin{equation}
\mathbf{\dot{u}}(\boldsymbol{\xi})=\mathbf{\Phi}(\boldsymbol{\xi})\cdot\mathbf{\dot{U}}(\boldsymbol{\xi})+\mathbf{\Theta}(\boldsymbol{\xi}):\mathbf{\Sigma}(\boldsymbol{\xi})
\end{equation}

where

\begin{equation}\label{APhi}
\Phi_{ij}=\omega^2\rho^0\left[a\zeta_i\zeta_j+b\delta_{ij}\right]
\end{equation}

and

\begin{equation}
\Theta_{ijp}=-\frac{1}{\omega\rho^0}\Phi_{ij}\zeta_p
\end{equation}

Since $\boldsymbol{\Theta}$ gets contracted with the symmetric tensor $\boldsymbol{\Sigma}$, it can be symmetrized about its last two indices. We make the following transformation,

\begin{eqnarray}\label{ATheta}
  \Theta_{ijp} & = & \frac{1}{2}\left[\Theta_{ijp}+\Theta_{ipj}\right]\\
   & = & -\frac{1}{2\omega\rho^0}\left[\Phi_{ij}\zeta_p+\Phi_{ip}\zeta_j\right]\nonumber \\
   & = & -\frac{\omega^2\rho^0}{2\omega\rho^0}\left[\frac{2\left[\lambda^0+\mu^0\right]}{\left[\omega^2\rho^0-\mu^0\boldsymbol{\zeta}^2\right]\left[\omega^2\rho^0-(\lambda^0+2\mu^0)\boldsymbol{\zeta}^2\right]}\zeta_i\zeta_j\zeta_p\right.\nonumber 
\\
   & + & \left.\frac{1}{\omega^2\rho^0-\mu^0\boldsymbol{\zeta}^2}\left[\delta_{ij}\zeta_p+\delta_{ip}\zeta_j\right]\right]\nonumber
\end{eqnarray}

The left hand side of Eq. (\ref{EStressXi}) can be expanded as,

\begin{eqnarray}\label{EFF}
 \lefteqn{-\mathbf{C}^0:[\boldsymbol{\zeta}\otimes(\boldsymbol{\zeta}\cdot\boldsymbol{\sigma})]_{sym}+\omega^2\rho^0\boldsymbol{\sigma} = F_{ijkl}\sigma_{kl}=}\\
   && \nonumber\left[-\lambda^0\delta_{ij}\zeta_k\zeta_l-\mu^0\frac{1}{2}\left[\zeta_i\delta_{jk}\zeta_l+\zeta_i\delta_{jl}\zeta_k+\zeta_j\delta_{ik}\zeta_l+\zeta_j\delta_{il}\zeta_k\right]+\omega^2\rho^0\delta_{ik}\delta_{jl}\right]\sigma_{kl}\\=
&& \nonumber\left[-\lambda^0g^{(2)}_{ijkl}-\mu^0g^{(1)}_{ijkl}+\omega^2\rho^0\delta_{ik}\delta_{jl}\right]\sigma_{kl}
\end{eqnarray}

where we define the following tensors,

\begin{equation}
g^{(1)}_{ijkl}=\frac{1}{2}\left[\zeta_i\delta_{jk}\zeta_l+\zeta_i\delta_{jl}\zeta_k+\zeta_j\delta_{ik}\zeta_l+\zeta_j\delta_{il}\zeta_k\right]
\end{equation}

\begin{equation}
g^{(2)}_{ijkl}=\delta_{ij}\zeta_k\zeta_l;\quad g^{(3)}_{ijkl}=\zeta_i\zeta_j\delta_{kl};\quad g^{(4)}_{ijkl}=\zeta_i\zeta_j\zeta_k\zeta_l
\end{equation}

and note the following relations,

\begin{equation}
\begin{array}{l}
\displaystyle g^{(1)}_{mnij}g^{(1)}_{ijkl}=2g^{(4)}_{mnkl}+\boldsymbol{\zeta}^2g^{(1)}_{mnkl}\\
\displaystyle g^{(1)}_{mnij}g^{(2)}_{ijkl}=2g^{(4)}_{mnkl}\\
\displaystyle g^{(1)}_{mnij}g^{(3)}_{ijkl}=2\boldsymbol{\zeta}^2g^{(3)}_{mnkl}\\
\displaystyle g^{(1)}_{mnij}g^{(4)}_{ijkl}=2\boldsymbol{\zeta}^2g^{(4)}_{mnkl}\\
\displaystyle g^{(2)}_{mnij}g^{(1)}_{ijkl}=2\boldsymbol{\zeta}^2g^{(2)}_{mnkl}\\
\displaystyle g^{(2)}_{mnij}g^{(2)}_{ijkl}=\boldsymbol{\zeta}^2g^{(2)}_{mnkl}\\
\displaystyle g^{(2)}_{mnij}g^{(3)}_{ijkl}=\boldsymbol{\zeta}^4\delta_{mn}\delta_{kl}\\
\displaystyle g^{(3)}_{mnij}g^{(1)}_{ijkl}=2g^{(4)}_{mnkl}\\
\displaystyle g^{(3)}_{mnij}g^{(2)}_{ijkl}=3g^{(4)}_{mnkl}\\
\displaystyle g^{(3)}_{mnij}g^{(3)}_{ijkl}=\boldsymbol{\zeta}^2g^{(3)}_{mnkl}\\
\displaystyle g^{(4)}_{mnij}g^{(1)}_{ijkl}=2\boldsymbol{\zeta}^2g^{(4)}_{mnkl}\\
\displaystyle g^{(4)}_{mnij}g^{(2)}_{ijkl}=\boldsymbol{\zeta}^2g^{(4)}_{mnkl}\\
\end{array} 
\end{equation}

To invert the tensor $\mathbf{F}$ in Eq. (\ref{EFF}) we seek a tensor \textbf{G} such that,

\begin{equation}
\mathbf{G}:\mathbf{F}=\mathbf{1}^{4s}
\end{equation}

Expressing $\mathbf{G}$ in terms of $\mathbf{g}^{(i)}$ and $\mathbf{1}^{4s}$ we have,

\begin{equation}
\mathbf{G}=a\mathbf{g}^{(1)}+b\mathbf{g}^{(2)}+c\mathbf{g}^{(3)}+d\mathbf{g}^{(4)}+e\mathbf{1}^{4s}
\end{equation}

Contracting with $\mathbf{F}$ we get the following equation,

\begin{eqnarray*}
\left[a(\omega^2\rho^0-\mu^0\boldsymbol{\zeta}^2)-e\mu^0\right]\mathbf{g}^{(1)}+\left[b(\omega^2\rho^0-(\lambda^0+2\mu^0)\boldsymbol{\zeta}^2)-e\lambda^0\right]\mathbf{g}^{(2)}\\
+\left[c\omega^2\rho^0\right]\mathbf{g}^{(3)}+\left[-a(2\lambda^0+2\mu^0)-c(3\lambda^0+2\mu^0)-d(\omega^2\rho^0-(\lambda^0+2\mu^0)\boldsymbol{\zeta}^2)\right]\mathbf{g}^{(4)}\\
+e\omega^2\rho^0\mathbf{1}^{4s}=\mathbf{1}^{4s}
\end{eqnarray*}

Solving the above equation we get the following values for the constants

\begin{equation}
\begin{array}{l}
\displaystyle a=\frac{\mu^0}{\omega^2\rho^0\left[\omega^2\rho^0-\mu^0\boldsymbol{\zeta}^2\right]}\\
\displaystyle b=\frac{\lambda^0}{\omega^2\rho^0\left[\omega^2\rho^0-(\lambda^0+2\mu^0)\boldsymbol{\zeta}^2\right]}\\
\displaystyle c=0\\
\displaystyle d=\frac{2\mu^0\left[\lambda^0+\mu^0\right]}{\omega^2\rho^0\left[\omega^2\rho^0-\mu^0\boldsymbol{\zeta}^2\right]\left[\omega^2\rho^0-(\lambda^0+2\mu^0)\boldsymbol{\zeta}^2\right]}\\
\displaystyle e=\frac{1}{\omega^2\rho^0}\\

\end{array} 
\end{equation}

Now Eq. (\ref{EStressXi}) can be written as,

\begin{equation}
\boldsymbol{{\sigma}}(\boldsymbol{\xi})=\mathbf{\Psi}(\boldsymbol{\xi})\cdot\mathbf{\dot{U}}(\boldsymbol{\xi})+\mathbf{\Gamma}(\boldsymbol{\xi}):\mathbf{\Sigma}(\boldsymbol{\xi})
\end{equation}

where,

\begin{eqnarray}\label{APsi}
  \Psi_{ijp} & = & -\frac{\omega\rho^0}{2}G_{ijkl}C^0_{klmn}\left[\zeta_m\delta_{np}+\zeta_n\delta_{mp}\right]\\
   & = & -\frac{\omega\rho^0}{2}\left[\left\{\frac{2\mu^0(\lambda^0+\mu^0)}{[\omega^2\rho^0-\mu^0\boldsymbol{\zeta}^2][\omega^2\rho^0-(\lambda^0+2\mu^0)\boldsymbol{\zeta}^2]}\right\}\zeta_i\zeta_j\zeta_p\right. \\
   & + & \left.\left\{\frac{\lambda^0}{\omega^2\rho^0-(\lambda^0+2\mu^0)\boldsymbol{\zeta}^2}\right\}\delta_{ij}\zeta_p+\left\{\frac{\mu^0}{\omega^2\rho^0-\mu^0\boldsymbol{\zeta}^2}\right\}\{\zeta_i\delta_{jp}+\zeta_j\delta_{ip}\}\right]\nonumber
\end{eqnarray}

and

\begin{eqnarray}\label{AGamma}
   \Gamma_{ijkl} & = & \omega^2\rho^0G_{ijkl}\\
   & = & \frac{\mu^0}{\omega^2\rho^0-\mu^0\boldsymbol{\zeta}^2}\frac{1}{2}\{\zeta_i\delta_{jk}\zeta_l+\zeta_i\delta_{jl}\zeta_k+\zeta_j\delta_{ik}\zeta_l+\zeta_j\delta_{il}\zeta_k\}  \\
   & + & \frac{\lambda^0}{\omega^2\rho^0-(\lambda^0+2\mu^0)\boldsymbol{\zeta}^2}\delta_{ij}\zeta_k\zeta_l\nonumber
\\
   & + & \frac{2\mu^0[\lambda^0+\mu^0]}{[\omega^2\rho^0-\mu^0\boldsymbol{\zeta}^2][\omega^2\rho^0-(\lambda^0+2\mu^0)\boldsymbol{\zeta}^2]}\zeta_i\zeta_j\zeta_k\zeta_l+\frac{1}{2}\left[\delta_{ik}\delta_{jl}+\delta_{il}\delta_{jk}\right]\nonumber
\end{eqnarray}

The components of tensors $\boldsymbol{\Phi}$, $\boldsymbol{\Theta}$, $\boldsymbol{\Gamma}$, and $\boldsymbol{\Psi}$ can be expressed in terms of the longitudinal wave speed $c_1=\sqrt{(\lambda^0+2\mu^0)/\rho^0}$ and the shear wave speed $c_1=\sqrt{\mu^0/\rho^0}$,

\begin{equation}
\Phi_{ij}=\omega^2\left[\frac{c_1^2-c_2^2}{\left[\omega^2-c_2^2\boldsymbol{\zeta}^2\right]\left[\omega^2-c_1^2\boldsymbol{\zeta}^2\right]}\zeta_i\zeta_j+\frac{1}{\left[\omega^2-c_2^2\boldsymbol{\zeta}^2\right]}\delta_{ij}\right]
\end{equation}

\begin{equation}
\Theta_{ijp}=-\frac{1}{\omega\rho^0}\Phi_{ij}\zeta_p
\end{equation}

\begin{eqnarray}
  \Gamma_{ijkl} & = & \frac{c_2^2}{\omega^2-c_2^2\boldsymbol{\zeta}^2}\frac{1}{2}\{\zeta_i\delta_{jk}\zeta_l+\zeta_i\delta_{jl}\zeta_k+\zeta_j\delta_{ik}\zeta_l+\zeta_j\delta_{il}\zeta_k\} \\
  & & {} +\frac{c_1^2-2c_2^2}{\omega^2-c_1^2\boldsymbol{\zeta}^2}\delta_{ij}\zeta_k\zeta_l\nonumber+\frac{2c_2^2[c_1^2-c_2^2]}{[\omega^2-c_2^2\boldsymbol{\zeta}^2][\omega^2-c_1^2\boldsymbol{\zeta}^2]}\zeta_i\zeta_j\zeta_k\zeta_l+\frac{1}{2}\left[\delta_{ik}\delta_{jl}+\delta_{il}\delta_{jk}\right]\nonumber
\end{eqnarray}

\begin{eqnarray}
  \Psi_{ijp} & = & -\omega\rho^0\left[\left\{\frac{2c_2^2(c_1^2-c_2^2)}{[\omega^2-c_2^2\boldsymbol{\zeta}^2][\omega^2-c_1^2\boldsymbol{\zeta}^2]}\right\}\zeta_i\zeta_j\zeta_p\right. \\
  & & {} \left.+\left\{\frac{c_1^2-2c_2^2}{\omega^2-c_1^2\boldsymbol{\zeta}^2}\right\}\delta_{ij}\zeta_p+\left\{\frac{c_2^2}{\omega^2-c_2^2\boldsymbol{\zeta}^2}\right\}\{\zeta_i\delta_{jp}+\zeta_j\delta_{ip}\}\right]\nonumber
\end{eqnarray}

To facilitate solution we convert the tensorial equations into their matrix form in the main text which are then used to prove some essential behaviour of the tensors. These proofs rely on transforming the tensors into forms where their inter-relationships are more transparent. To that end we transform the tensor $\boldsymbol{\Psi}$ to,

\begin{equation}
\hat{\Psi}_{mnp}=D^0_{mnij}\Psi_{ijp}
\end{equation}

where $\mathbf{D}^0$ is the compliance tensor given by

\begin{equation}
D^0_{mnij}=\frac{-\lambda^0}{2\mu^0(3\lambda^0+2\mu^0)}\delta_{mn}\delta_{ij}+\frac{1}{4\mu^0}(\delta_{mi}\delta_{nj}+\delta_{mj}\delta_{ni})
\end{equation}

The final expression for $\hat{\boldsymbol{}\Psi}$ is given below,

\begin{eqnarray}\label{APsiHat}
  \hat{\Psi}_{mnp} & = & D^0_{mnij}\Psi_{ijp}\\
   & = & -\frac{\omega}{2}\left[\frac{2\left[c_1^2-c_2^2\right]}{\left[\omega^2-c_2^2\boldsymbol{\zeta}^2\right]\left[\omega^2-c_1^2\boldsymbol{\zeta}^2\right]}\zeta_m\zeta_n\zeta_p\right.\nonumber 
\\
   & + & \left.\frac{1}{\omega^2-c_2^2\boldsymbol{\zeta}^2}\left[\zeta_m\delta_{np}+\zeta_n\delta_{mp}\right]\right]\nonumber
\end{eqnarray}

If we now define $\hat{\boldsymbol{\Theta}}$ as,

\begin{equation}\label{AThetaHat}
\hat{\Theta}_{mnp}=\rho^0\Theta_{mnp}
\end{equation}

we find from Eqs. (\ref{ATheta}, \ref{APsiHat}) that,

\begin{equation}
\hat{\Psi}_{mnp}=\hat{\Theta}_{pmn}
\end{equation}

We also define $\hat{\boldsymbol{\Phi}}$ as,

\begin{equation}
\hat{\Phi}_{ij}=\rho^0{\Phi}_{ij}
\end{equation}

and $\hat{\boldsymbol{\Gamma}}$ as,

\begin{eqnarray}\label{AGammaHat}
  \hat{\Gamma}_{mnkl} & = & D^0_{mnij}\Gamma_{ijkl}\\
   & = & \frac{1}{\rho^0}\left[\frac{1}{4(\omega^2-c_2^2\boldsymbol{\zeta}^2)}\{\zeta_m\delta_{nk}\zeta_l+\zeta_m\delta_{nl}\zeta_k+\zeta_n\delta_{mk}\zeta_l+\zeta_n\delta_{ml}\zeta_k\}\right.
\nonumber \\
   & + & 
\left. \frac{-(c_1^2-2c_2^2)}{2c_2^2(3c_1^2-4c_2^2)}\delta_{mn}\delta_{kl}+\frac{c_1^2-c_2^2}{[\omega^2-c_2^2\boldsymbol{\zeta}^2][\omega^2-c_1^2\boldsymbol{\zeta}^2]}\zeta_m\zeta_n\zeta_k\zeta_l\right.
\nonumber \\
   & + & 
\left.\frac{1}{4c_2^2}\{\delta_{mk}\delta_{nl}+\delta_{ml}\delta_{nk}\}\right]\nonumber
\end{eqnarray}

We notice that transformed as above, $\hat{\boldsymbol{\Gamma}}$ has the symmetries,

\begin{equation}
\hat{\Gamma}_{mnkl}=\hat{\Gamma}_{nmkl}=\hat{\Gamma}_{mnlk}=\hat{\Gamma}_{klmn}
\end{equation}
\end{document}